\documentclass[12pt]{article}
\usepackage{amssymb}
\begin{document}
\thispagestyle{empty}
\newcommand{\be}{\begin{equation}}
\newcommand{\ee}{\end{equation}}
\newcommand{\sect}[1]{\setcounter{equation}{0}\section{#1}}
\newcommand{\vs}[1]{\rule[- #1 mm]{0mm}{#1 mm}}
\newcommand{\hs}[1]{\hspace{#1mm}}
\newcommand{\mb}[1]{\hs{5}\mbox{#1}\hs{5}}
\newcommand{\bea}{\begin{eqnarray}}
\newcommand{\eea}{\end{eqnarray}}
\newcommand{\wt}[1]{\widetilde{#1}}
\newcommand{\ux}[1]{\underline{#1}}
\newcommand{\ov}[1]{\overline{#1}}
\newcommand{\sm}[2]{\frac{\mbox{\footnotesize #1}\vs{-2}}
           {\vs{-2}\mbox{\footnotesize #2}}}
\newcommand{\prt}{\partial}
\newcommand{\eps}{\epsilon}\newcommand{\p}[1]{(\ref{#1})}
\newcommand{\R}{\mbox{\rule{0.2mm}{2.8mm}\hspace{-1.5mm} R}}
\newcommand{\Z}{Z\hspace{-2mm}Z}
\newcommand{\cd}{{\cal D}}
\newcommand{\cg}{{\cal G}}
\newcommand{\ck}{{\cal K}}
\newcommand{\cw}{{\cal W}}
\newcommand{\vj}{\vec{J}}
\newcommand{\vl}{\vec{\lambda}}
\newcommand{\vz}{\vec{\sigma}}
\newcommand{\vt}{\vec{\tau}}
\newcommand{\poiss}{\stackrel{\otimes}{,}}
\newcommand{\tx}{\theta_{12}}
\newcommand{\tb}{\overline{\theta}_{12}}
\newcommand{\zw}{{1\over z_{12}}}
\newcommand{\sqp}{{(1 + i\sqrt{3})\over 2}}
\newcommand{\sqm}{{(1 - i\sqrt{3})\over 2}}
\newcommand{\NP}[1]{Nucl.\ Phys.\ {\bf #1}}
\newcommand{\PLB}[1]{Phys.\ Lett.\ {B \bf #1}}
\newcommand{\PLA}[1]{Phys.\ Lett.\ {A \bf #1}}
\newcommand{\NC}[1]{Nuovo Cimento {\bf #1}}
\newcommand{\CMP}[1]{Commun.\ Math.\ Phys.\ {\bf #1}}
\newcommand{\PR}[1]{Phys.\ Rev.\ {\bf #1}}
\newcommand{\PRL}[1]{Phys.\ Rev.\ Lett.\ {\bf #1}}
\newcommand{\MPL}[1]{Mod.\ Phys.\ Lett.\ {\bf #1}}
\newcommand{\BLMS}[1]{Bull.\ London Math.\ Soc.\ {\bf #1}}
\newcommand{\IJMP}[1]{Int.\ J.\ Mod.\ Phys.\ {\bf #1}}
\newcommand{\JMP}[1]{Jour.\ Math.\ Phys.\ {\bf #1}}
\newcommand{\LMP}[1]{Lett.\ Math.\ Phys.\ {\bf #1}}
\newpage
\setcounter{page}{0} \pagestyle{empty} \vs{12}
\begin{center}
{\Large {\bf On a Division Algebra Classification of
Constrained Generalized Supersymmetries.}}\\ {\quad}\\

\vs{10} {\large Francesco Toppan} ~\\ \quad
\\
 {\large{\em CCP - CBPF}}\\{\em Rua Dr. Xavier Sigaud
150, cep 22290-180 Rio de Janeiro (RJ)}\\{\em Brazil}\\

\end{center}
{\quad}\\ \centerline{ {\bf Abstract}}

\vs{6}

In this talk we present a division-algebra classification of the generalized supersymmetries admitting bosonic tensorial central charges. 
We show that for complex and quaternionic supersymmetries a whole class
of compatible division-algebra constraints can be imposed. Possible applications to $M$-theory related dynamical systems are briefly
mentioned.
\vs{6}
\begin{center}

\end{center}
\vfill
{Invited plenary talk at the IV International Conference on Mathematical Methods in Physics, August 2004, Rio de Janeiro, Brazil. Published in PoS WC2004/047.}\\
{{\em E-mail:}\quad {$toppan@cbpf.br$}}
\\
\pagestyle{plain}
\renewcommand{\thefootnote}{\arabic{footnote}}

\section{Introduction}
In this talk we present a work in progress, based on several papers of the author and his
coworkers, concerning the division algebra classification of the generalized supersymmetries and of their consistent constraints.\par
This is an extremely important issue. We recall in fact that
in the seventies
the H\L S scheme \cite{hls} was a cornerstone providing the supersymmetric
extension of the Coleman-Mandula no-go theorem. 
However, in the eighties \cite{daf} and especially in the nineties, the generalized space-time supersymmetries admitting bosonic tensorial central charges and going
beyond the H\L S scheme found recognition \cite{{agit},{tow}}
in association with the dynamics of extended objects like branes (see \cite{{gs},{ste}}).
The eleven-dimensional $M$-algebra underlying the $M$-theory as a possible ``Theory 
Of Everything" (TOE), admitting $32$-real component spinors and maximal number ($=528$) of
saturated bosonic generators \cite{{agit},{tow}} falls into this class of generalized supersymmetries.
This is the reason why a lot of attention has been recently devoted to the problem of classifying generalized supersymmetries, see e.g. \cite{fer} and \cite{acdp}.
A step towards this classification was provided in \cite{top}. Based on the available classification
of Clifford algebras and spinors in terms of division algebras \cite{{abs},{por},{oku}}, 
it was there shown that, in the complex and quaternionic cases, a division-algebra compatible
constraint,
leading to the two big classes of hermitian and holomorphic generalized supersymmetries,
could be consistently imposed. In this talk we review the main ingredients entering
the mathematical classification of generalized supersymmetries and present new results
\cite{kuto} on the classification of the constrained generalized supersymmetries.\par
The present paper is so conceived. In order to make it self-consistent, at first the division-algebra classification of Clifford algebras and fundamental spinors is recalled.
The notion of ``maximal Clifford algebras", essential for later developments, is introduced.
It is explained how to recover all real, complex and quaternionic realizations in any given
space-time from the set of fundamental maximal Clifford algebras which can be iteratively constructed. In the following, the notion of generalized supersymmetry is introduced in association with their division algebra properties.
It is further explained how to implement various division algebra-compatible constraints,
as well as their combinations. This amounts to introduce hermitian versus holomorphic constraints in the complex and quaternionic cases, as well as reality conditions implemented
on bosonic generators.
Some concrete examples of these division-algebra compatible constrained generalized supersymmetries are explicitly constructed. A series of tables with the main
ingredients of the classification are presented. Finally, in the Conclusions, we will briefly mention the possible physical applications to supersymmetric dynamical systems (and their relation with the $M$-theory) of the above construction.

\section{Basic notions: division algebras, Clifford algebras and fundamental spinors.}

The four division algebras of real (${\bf R}$) and complex (${\bf C}$) numbers, quaternions (${\bf H}$)
and octonions 
(${\bf O}$) possess respectively $0$, $1$, $3$ and $7$ imaginary elements $e_i$ satisfying the relations
\begin{eqnarray}
e_i\cdot e_j &=& -\delta_{ij} + C_{ijk} e_{k},
\label{octonrel}
\end{eqnarray}
($i,j,k$ are restricted to take the value $1$ in the complex case, $1,2,3$ in the quaternionic case and
$1,2,\ldots , 7$ in the octonionic case; furthermore, the sum over repeated indices is understood).\par
$C_{ijk}$ are the totally antisymmetric division-algebra structure constants. The octonionic division
algebra is the maximal, since quaternions, complex and real numbers can be obtained as its restriction.
The totally antisymmetric octonionic structure constants can be expressed as
\begin{eqnarray}
&C_{123}=C_{147}=C_{165}=C_{246}=C_{257}=C_{354}=C_{367}=1&
\end{eqnarray}
(and vanishing otherwise).
\par
The octonions are the only non-associative, however alternative (see \cite{bae}), division algebra.\par
For our later purposes it is of particular importance the notion of division-algebra principal conjugation.
Any element $X$ in the given division algebra can be expressed through the sum
\begin{eqnarray}
X&=&x_0 + x_ie_i ,
\end{eqnarray}
where $x_0$ and $x_i$ are real, the summation over repeated indices is understood and the positive integral
$i$ are restricted up to $1$, $3$ and $7$ in the ${\bf C}$, ${\bf H}$ and ${\bf O}$ cases respectively.
The principal conjugate $X^\ast$ of $X$ is defined to be
\begin{eqnarray}\label{conjug}
X^\ast&=&x_0 - x_ie_i .
\end{eqnarray}
It allows introducing the division-algebra norm through the product $X^\ast X$. The 
normed-one restrictions $X^\ast X =1$ select the three parallelizable spheres $S^1$, $S^3$ and $S^7$
in association with ${\bf C}$, ${\bf H}$ and ${\bf O}$ respectively.
\par
For what concerns the main properties of Clifford algebras and their relation with the associative division algebras ${\bf R}, ${\bf C}, ${\bf H}$ it is 
convenient to follow \cite{oku} and \cite{crt1}. 
\par The most general irreducible {\em real}
matrix representations of the Clifford algebra
\begin{eqnarray}
\Gamma^\mu\Gamma^\nu+\Gamma^\nu\Gamma^\mu &=& 2\eta^{\mu\nu},
\label{cliff}
\end{eqnarray}
with $\eta^{\mu\nu}$ being a diagonal matrix of $(p,q)$ signature
(i.e. $p$ positive, $+1$, and $q$ negative, $-1$, diagonal
entries)\footnote{Throughout this paper it will be understood that the positive eigenvalues are associated
with space-like directions, the negative ones with time-like directions.}
can be classified according to the property of the most
general $S$ matrix commuting with all the $\Gamma$'s ($\relax
[S,\Gamma^\mu ] =0$ for all $\mu$). If the most general $S$ is a
multiple of the identity, we get the normal (${\bf R}$) case.
Otherwise, $S$ can be the sum of two matrices, the second one
multiple of the square root of $-1$ (this is the almost complex,
${\bf C}$ case) or the linear combination of $4$ matrices closing
the quaternionic algebra (this is the ${\bf H}$ case).
\par
For our purposes the division-algebra character of the Clifford irreps can be understood as follows. In the
${\bf R}$-case the matrices realizing the irrep have necessarily real entries,
in the ${\bf C}$-case matrices with complex entries can be used, while in the ${\bf H}$-case
the matrices can be realized with quaternionic entries.\par
Let us see how this works in a simple example. 
Let us take the ${\bf H}$-type $C(0,3)$ Clifford algebra. It
can be realized by associating the three Euclidean gamma matrices with the three 
imaginary quaternions $e_{i}$. The reason for that lies on the fact that the antisimmetry
of the $C_{ijk}$ (\ref{octonrel}) structure constants make the anticommutators
$e_ie_j+e_je_i$ satisfy the relation
\begin{eqnarray}\label{quat}
e_ie_j+e_je_i&=& -\delta_{ij},
\end{eqnarray}
reproducing the three dimensional Euclidean Clifford algebra (\ref{cliff})
with negative signs.
\par
It is worth mentioning that in the given signatures $p-q~mod~8 = 0,4,6,7$,
without loss of generality, the $\Gamma^\mu$ matrices can be
chosen block-antidiagonal (generalized Weyl-type matrices), i.e.
of the form
\begin{eqnarray}
\Gamma^\mu &=&\left( \begin{array}{cc}
  0 & \sigma^\mu \\
  {\tilde\sigma}^\mu & 0
\end{array}\right)\label{weyl}
\end{eqnarray}
Since the generalized Lorentz algebra can be recovered from the algebra of the
commutators $\relax \Sigma^{[\mu\nu]} = [\Gamma^\mu,\Gamma^\nu]$, in those 
particular signatures the matrices $\Sigma^{[\mu\nu]}$ are of block-diagonal type
and it is therefore possible to introduce Weyl-projected spinors, whose number of components is half of the
size of the corresponding $\Gamma$-matrices (this notion of Weyl spinors,
which is convenient for our purposes,
has been introduced in \cite{crt1}).\par
In \cite{crt1} the representatives of all Clifford irreps in any given space-time were systematically
constructed with the help of two recursive algorithms (producing $D+2$-dimensional
Clifford irreps from $D$-dimensional spacetime Clifford irreps),
to be applied to the solutions (such as (\ref{quat})) of the equation (\ref{cliff}) in terms of imaginary elements
of a division algebra (see \cite{crt1} for detail).  
\par
Let us briefly comment about the octonionic realization of the (\ref{cliff}) relation,
through matrices admitting octonionic entries. Since the octonions are non-associative, this realization presents peculiar features. In \cite{lt1} and \cite{lt2} it was shown how it could
be associated with an octonionic version of the $M$ algebra and its associated superconformal algebra. Throughout this paper we will limit ourselves to consider only standard, associative,
Clifford algebras representations.
\par
Fundamental spinors carry a representation of the generalized Lorentz group with a minimal number of real components in association with the maximal, compatible, allowed division-algebra structure (they can be thought as column vectors with entries in the given
division algebra).\par
It is worth reminding that the division-algebra character of fundamental spinors does not necessarily (depending on the given space-time) coincide with the division-algebra type of the corresponding Clifford irreps (this mismatch lies on the fact that in some 
given spacetimes the fundamental spinors are of Weyl type).
In different space-times parametrized by $\rho= s-t\quad mod\quad 8$,
fundamental spinors can accommodate for $\rho=2,3$ a larger
division-algebra structure than the corresponding Clifford irreps. Conversely, for $\rho= 6,7$, the Clifford irreps
accommodate a larger division-algebra structure than the corresponding spinors.

\section{Maximal Clifford algebras and their reductions.}

An extremely useful notion is that of ``maximal Clifford algebra" (see \cite{crt1}).
``Maximal Clifford algebras" correspond
to the Clifford irreps which can accommodate the maximal number of Gamma matrices for the corresponding size of the matrices. Stated otherwise,for any given spacetime, its Clifford irrep can be obtained from its associated maximal Clifford algebras. Non-maximal Clifford algebras are simply recovered after deleting a certain number of Gamma matrices from a given maximal one (a procedure which parallels the
dimensional reduction). \par
The knowledge of maximal Clifford algebras (which can be obtained with the lifting algorithms of \cite{crt1}) allows us to reconstruct the full set of Clifford irreps in any given space-time.\par
The maximal Clifford irreps exist iff $(p-q)=1,5\quad mod\quad 8$.
The $(p-q)=1\quad mod\quad 8$ condition corresponds to a real case, while the
$p-q =5\quad mod\quad 8$ condition corresponds to a quaternionic case. 
The non-maximal Clifford algebras, given by $p-q \neq 1,5\quad mod \quad 8$,
can be recovered with the procedure illustrated by the following table \cite{kuto}, specifying
real, complex and quaternionic
Clifford irreps (denoted as $\Gamma$) and Clifford representations (not necessarily irrep, denoted as $\Psi$) supporting fundamental spinors:
{\small
 { {{\begin{eqnarray}&\label{maxclifford}
\begin{tabular}{|l|l|l|}\hline
$$&${1\quad mod\quad 8\quad ({\bf R})}$&${5\quad mod\quad 8 \quad ({\bf H})}$
\\ \hline

$0\quad mod\quad 8$&$\Gamma,\Psi: (p,q)\stackrel{W}{\rightarrow} (p-1,q)$&$$\\ \hline

$4\quad mod \quad 8$&$ $&$\Gamma,\Psi: (p,q)\stackrel{W}{\rightarrow} (p-1,q)$\\ \hline

$2\quad mod\quad 8$&$\Gamma: (p,q)\stackrel{}{\rightarrow} (p,q-1)$&$
\Psi:(p,q)\stackrel{\ast}{\rightarrow} (p-2,q)\stackrel{W}{\rightarrow} (p-3,q)$\\ \hline

 $3\quad mod\quad 8$&$
$&$\begin{tabular}{l}$\Gamma: (p,q)\stackrel{\ast}{\rightarrow} (p-2,q)$\\
$\Psi: (p,q)\stackrel{W}{\rightarrow} (p-2,q)$
\end{tabular}$\\ \hline

$6\quad mod\quad 8$&$$&$
\begin{tabular}{l}$
\Gamma: (p,q){\rightarrow} (p,q-1)$\\$
\Psi: (p,q)\stackrel{\ast}{\rightarrow} (p,q-2)\stackrel{W}{\rightarrow} (p-1,q-2)
$\end{tabular}$\\ \hline

 $7\quad mod\quad 8$&$\Psi: (p,q)\stackrel{W}{\rightarrow} (p-2,q)
$&$\Gamma: (p,q)\stackrel{\ast}{\rightarrow} (p,q-2)$\\ \hline

\end{tabular}&\nonumber\\&&\end{eqnarray}}} }}
Some remarks are in order. The real case is shown in the second column, while both
the complex and the quaternionic cases are recovered from the third column. The arrows denote
which gamma matrices (either space-like or time-like) and how many of them have to be deleted
from the corresponding maximal Clifford algebra. The ``$W$" symbol above an arrow specifies whether the Weyl projection is required in order to produce fundamental spinors,
while the ``$\ast$" symbol above an arrow denotes a reduction to the complex case.\par
The $(p,q)\stackrel{\ast}{\rightarrow} (p-2,q)$ reduction can only be performed under the condition
$p\geq 3$, see \cite{kuto} for details. Similarly, the $(p,q)\stackrel{\ast}{\rightarrow} (p,q-2)$ 
reduction requires $q\geq 3$ (all cases of physical interest enter the above table, the remaining few exceptional cases can be treated separately). 

\section{On generalized supersymmetries.}

Let us introduce now the notion of generalized supersymmetries as an extension and generalization of
the standard supertranslation algebra (in some cases, like the $F$-algebra presentation
in a $(10,2)$ spacetime of the $M$-algebra \cite{top}, the bosonic sector admits no translation at all, but still it is convenient to refer to generalized supersymmetries as ``generalized supertranslations").  Generalized supertranslations can be used as building
blocks to construct superconformal algebras (by simply taking two separate copies of generalized supertranslations and then imposing the closure of the super-Jacobi identities
on all generators, \cite{lt1}). Once obtained a generalized superconformal algebra, generalized superPoincar\'e algebras admitting, besides the generalized supertranslations,
also the generalized Lorentz generators, can be recovered through an Inon\"u-Wigner contraction
procedure. Throughout this paper we will focus just on the building blocks, namely the generalized supertranslations.\par 
At first we need to introduce two matrices, denoted as $A$ and $C$\cite{kt}, related with,
respectively, the hermitian
conjugation and transposition acting on Gamma matrices.
$A$ plays the role of the time-like $\Gamma^0$ matrix in
the Minkowskian space-time and is used to introduce barred spinors. $C$, on the other hand,
is the charge conjugation matrix. Up to an overall sign, in a generic $(p,q)$ space-time, $A$ and $C$
are given by the products of all the time-like and, respectively, all the symmetric (or antisymmetric) Gamma-matrices (depending on the given space-time there are at most two charge conjugations matrices,
$C_S$, $C_A$, given by the product of all symmetric and all antisymmetric gamma matrices).
For our purposes the importance of $A$ and the charge conjugation matrix $C$ lies on the fact that, in a 
$D$-dimensional space-time ($D=p+q$) spanned by $d\times d$ Gamma matrices, they allow to construct a basis for $d\times d$ (anti)hermitian and (anti)symmetric matrices, respectively.
The
$\left( \begin{array}{c}
  D\\
  k
\end{array}\right)$ antisymmetrized products of $k$ Gamma matrices
$A{\Gamma}^{[\mu_1 \ldots \mu_k]}$ are all hermitian or all antihermitian, depending on the value 
of $k\leq D$. Similarly, the antisymmetrized products $C{\Gamma}^{[\mu_1 \ldots \mu_k]}$ are all 
symmetric or all antisymmetric.\par
A generalized supersymmetry algebra involving $n$-component real spinors $Q_a$ is given by the anticommutators
\begin{eqnarray}\label{Mgen}
    \left\{ Q_a, Q_b \right\} & = & {\cal Z}_{ab},
\end{eqnarray}
where the matrix ${\cal Z} $ appearing in the r.h.s. is the most general $n\times n$
symmetric matrix with total number of $\frac{n(n+1)}{2}$ components. For any given space-time we
can easily compute its associated decomposition
in terms of the antisymmetrized products of $k$-Gamma matrices, namely
\begin{eqnarray}
{\cal Z}_{ab} &=& \sum_k(C\Gamma_{[\mu_1\ldots\mu_k]})_{ab}Z^{[\mu_1\ldots \mu_k]},
\end{eqnarray}
where the values $k$ entering the sum in the r.h.s. are restricted by the symmetry requirement for the 
$a\leftrightarrow b$ exchange
and are specific for the given spacetime. The coefficients $Z^{[\mu_1\ldots \mu_k]}$ are the rank-$k$ abelian tensorial central charges.  \par
In the case of Weyl projected spinors ${\widetilde Q}_a$ the r.h.s. has to be reconstructed with the help of a projection operator which selects the upper left block in a $2\times 2$ block decomposition.
Specifically, if ${\cal Z}$ is a matrix decomposed in $2\times 2$ blocks as
${\cal Z} =\left( \begin{array}{cc}
  {\cal Z}_1&{\cal Z}_2\\
  {\cal Z}_3&{\cal Z}_4
\end{array}\right)$, we can define 
\begin{eqnarray}\label{pweyl}
P({\cal Z}) &\equiv & {\cal Z}_1.
\end{eqnarray} 
The generalized supersymmetry algebra in the Weyl case can be expressed through
\begin{eqnarray}\label{weylproj}
    \left\{ {\widetilde Q}_a, {\widetilde Q}_b \right\} & = & P({\cal Z})_{ab}.
\end{eqnarray}
A complex (quaternionic) generalized supersymmetry algebra is expressed in terms of complex (quaternionic) spinors $Q_a$ and their conjugate ${Q^\ast}_{\dot a}$.
The most general (with a saturated r.h.s.) algebra is in this case given by 
\begin{eqnarray}\label{Mhol}
    \left\{ Q_a, Q_b \right\} =  {\cal P}_{ab}\quad &,& \quad \left\{ {Q^\ast}_{\dot a}, {Q^\ast}_{\dot b} \right\} =  {{\cal P}^\ast}_{\dot{a}\dot{b}},
\end{eqnarray}
together with
\begin{eqnarray}\label{Mher}
\left\{ Q_a, {Q^\ast}_{\dot b} \right\} &=&  {\cal R}_{{a}\dot{b}},
\end{eqnarray}
where the matrix ${\cal P}_{ab}$ (${{\cal P}^\ast}_{\dot{a}\dot{b}}$ is its conjugate and does not contain new degrees of freedom) is symmetric,
while ${\cal R}_{{a}\dot{b}}$ is hermitian.\par
The maximal number of allowed components in the r.h.s. is given, for complex fundamental spinors
with $n$ complex components, by $n(n+1)$ (real) bosonic components entering the symmetric $n\times n$ complex matrix ${\cal P}_{ab}$ 
plus $n^2$  (real) bosonic components entering the hermitian $n\times n$ complex matrix 
${\cal R}_{{a}\dot{b}}$.\par
A Weyl projection similar to (\ref{weylproj}) can be applied for complex and quaternionic  spinors as well.

\section{Real generalized supersymmetries.}

In this section we present a series of tables, taken from \cite{top} and \cite{kuto},
listing the main properties of real generalized supersymmetries.
\par
It is convenient to symbolically denoted as ``${M}_k$" the 
space of $\left( \begin{array}{c}
  D  \\ k
\end{array}\right)$-component,
totally antisymmetric rank-$k$ tensors of a $D$-dimensional 
spacetime, associated to the
basis of the symmetric
matrices $C\Gamma^{[\mu_1\ldots\mu_k]}$.\par
In the case of generalized real supersymmetries, depending on the 
dimensionality $D$ of the space-time (and independently from its signature, provided that
the spinors admit the same minimal number of components), the bosonic sector, together with its
number of bosonic components, is reported in the following table. 
Since maximal Clifford algebras are odd-dimensional, without loss of generality 
only odd dimensions $D$ enter the table below
{ {{\begin{eqnarray}&\label{realodd}
\begin{tabular}{|c|c|c|}\hline
spacetime&bosonic sectors&bosonic components\\ \hline
$D=1$&${M}_0$& $1$\\ \hline
$D=3$&${M}_1$& $3$\\ \hline
$D=5$&${{M}}_2$&$10$\\ \hline
$D=7$&${M}_0+{M}_3$&$1+35=36$\\ \hline
$D=9$&${M}_0+{M}_1+{M}_4$&$1+9+126=136$\\ \hline
$D=11$&${M}_1+{M}_2+{M}_5$&$11+55+462=528$\\ \hline
$D=13$&${M}_2+{M}_3+{M}_6$&$78+286+1716=2080$\\ \hline
\end{tabular}&\end{eqnarray}}} }  
Generalized supersymmetries 
in even dimensional spacetime can be obtained from the previous list  
via a dimensional reduction (by erasing some Gamma matrices, as explained in Section {\bf 3}).
We obtain that the dimensional reduction $D\rightarrow D-1$ corresponding to the signature passage $(p,q)\rightarrow (p,q-1)$ (here $D=p+q$) is expressed through
{\small{
{ {{\begin{eqnarray}&\label{realeven}
\begin{tabular}{|l|l|l|}\hline
spacetime&bosonic sectors&bosonic components\\ \hline
$D=3$&${M}_1\rightarrow \overline{M}_1+\overline{M}_0$& $3=2+1$\\ \hline
$D=5$&${{M}}_2\rightarrow \overline{M}_2+\overline{M}_1$&$10=6+4$\\ \hline
$D=7$&${M}_0+{M}_3\rightarrow \overline{M}_0+\overline{M}_3+
\overline{M}_2$&$36=1+20+15$\\ \hline
$D=9$&${M}_0+{M}_1+{M}_4\rightarrow 2\times\overline{M}_0+
\overline{M}_1+\overline{M}_4+
\overline{M}_3$&$136=2+8+70+56$\\ \hline
$D=11$&${M}_1+{M}_2+{M}_5\rightarrow \overline{M}_0+\overline{M}2\times\overline{M}_1+\overline{M}_2+
\overline{M}_4+\overline{M}_5$&$528=1+20+45+210+252$\\ \hline
$D=13$&${M}_2+{M}_3+{M}_6\rightarrow\overline{M}_1+2\times\overline{M}_2+\overline{M}_3
+\overline{M}_5+\overline{M}_6$&$2080=12+2\times 66+220+792+924$\\ \hline
\end{tabular}&\nonumber\\
&&\end{eqnarray}}} }  
}}
The overlined quantities $\overline{M}_k$ are 
referred to the totally antisymmetric $k$-tensors in the $D-1$-dimensional 
spacetime.\par
It is also convenient to illustrate the dimensional reduction leading from 
the $(p,q)\rightarrow (p-1,q)$ spacetime. The difference w.r.t. the previous case 
lies on the fact that now the
$(p-1,q)$ spacetime is of Weyl type (confront the discussion 
in Section {\bf 2}). Only the subclass of totally antisymmetric bosonic $k$-tensors 
entering the upper left diagonal block will survive from the Weyl projection and enter
the generalized supersymmetry. The corresponding symbols are marked 
in boldface $({\bf M}_k)$ in the table below, corresponding to the even-dimensional
Weyl case
{{
{ {{\begin{eqnarray}&\label{realevenweyl}
\begin{tabular}{|l|l|l|}\hline
spacetime&bosonic sectors&bosonic components\\ \hline
$D=2$&${M}_0+\frac{1}{2}{\bf{M}}_1$& $1$\\ \hline
$D=4$&$\frac{1}{2}{\bf{M}}_2+{M}_1$&$3$\\ \hline
$D=6$&${M}_0+\frac{1}{2}{\bf{M}}_3+{M}_2$&$10$\\ \hline
$D=8$&${\bf{M}}_0+{M}_1+{M}_3+\frac{1}{2}{\bf{M}}_4$&$36=1+35$\\ \hline
$D=10$&${M}_0+{\bf{M}}_1+{M}_2+{M}_4+\frac{1}{2}{\bf{M}}_5$&$136=10+126$\\ \hline
$D=12$&${M}_1+{\bf{M}}_2+{M}_3+{M}_5+2+\frac{1}{2}{\bf{M}}_6$&$528=66+462$\\ \hline
\end{tabular}&\nonumber\\
&&\end{eqnarray}}} }  
}}
In the above table the factor $\frac{1}{2}$ has been inserted to remind that ${\bf M}_{\frac{D}{2}}$ is self-dual, so that its total number of components has to be halved
in order to fulfill the selfduality condition.\par

\section{Constrained complex generalized supersymmetries}

Two big classes of subalgebras, respecting the Lorentz-covariance, can be obtained from (\ref{Mhol}) and (\ref{Mher}) in both the complex and quaternionic cases, by setting identically equal to zero either ${\cal P}$ or ${\cal R}$, namely
assuming that either
\begin{eqnarray}
&{\cal P}_{ab}\equiv   {{\cal P}^\ast}_{\dot{a}\dot{b}}\equiv 0,&\end{eqnarray} 
so that the only bosonic degrees
of freedom enter the hermitian matrix ${{\cal R}}_{{a}\dot{b}}$ or, conversely,
that
\begin{eqnarray}&{{\cal R}}_{{a}\dot{b}}\equiv 0,&\end{eqnarray} 
so that the only bosonic degrees of freedom enter 
${\cal P}_{ab}$ and its conjugate matrix ${{\cal P}^\ast}_{\dot{a}\dot{b}}$.\par
Following \cite{top} we will refer to the (complex or quaternionic) generalized supersymmetries satisfying
the first constraint as ``hermitian" generalized supersymmetries, while the
(complex or quaternionic) generalized supersymmetries satisfying the second constraint will be referred to
as ``holomorphic" generalized supersymmetries. This distinction finds application in physics.
It was proven in \cite{lt3} that the analytical continuation of the $M$-algebra can be carried out to the Euclidean, the corresponding Euclidean algebra being a complex holomorphic supersymmetry.
\par
Further refinements in the classification of division algebra constrained generalized supersymmetries can be produced by allowing a reality (or imaginary) constraint on the bosonic matrices ${\cal P}$ and ${\cal R}$.  It is convenient to illustrate it by discussing, at first, some specific examples
of interest, for later producing general results.\par
Let us start describing the generalized supersymmetries associated with the
$(4,1)$ space-time. Its fundamental spinors are quaternionic 
and admit $8$ real components. We are in the position to classify all real and complex
saturated generalized supersymmetries associated to this spacetime (the quaternionic supersymmetries are introduced in the next section).
There are seven separated cases that we are able to consider, depending on the mixed conditions
(holomorphicity, hermiticity, reality or imaginary constraint on ${\cal P}$ and ${\cal R}$) 
that can be imposed. The complete class of constrained generalized
supersymmetries can be given as follows:\par
{\em i)} Real generalized supersymmetry with $36$ bosonic components. This real generalized
supersymmetry can also be expressed in the complex spinor formalism, the $36$ bosonic components being recovered from $36=20+16$, the sum (in the real counting) of the holomorphic and hermitian sectors of the bosonic r.h.s.,\par
{\em ii)} A constrained complex supersymmetry obtained by imposing a reality condition on ${\cal R}$ alone. The total number of bosonic components in this case is $30$, \par
{\em iii)} The constraint arising by impoing either a reality condition on ${\cal P}$ or,
altenatively, an imaginary condition on ${\cal R}$. The total number of bosonic components is 
$26$,\par
{\em iv)}
The holomorphically constrained complex generalized supersymmetry with $20$ bosonic
components in the real counting (alternatively described by a reality condition on both ${\cal P}$ and ${\cal R}$,\par
{\em v)} The hermitian complex generalized supersymmetry with $16$ bosonic components (real counting) (alternatively described by a reality condition on ${\cal P}$ and an imaginary condition on ${\cal R}$),\par
{\em vi)} The holomorphically (or hermitian) constrained complex generalized supersymmetry with reality condition on the bosonic r.h.s., leading to $\frac{1}{2}\times 20 = 10$ bosonic components and, finally,\par
{\em vii)} the hermitian supersymmetry with an imaginary constraint on the bosonic sector, leading to $6$ bosonic components.\par
The generalized supersymmetries for $(4,1)$ allow us to immediately construct the generalized
supersymmetries in the standard Minkowskian $(3,1)$ space-time, which can be obtained as a 
Weyl-type dimensional reduction from $(4,1)$, see table (\ref{maxclifford}).  The corresponding generalized supersymmetries in this case admit a total number of bosonic generators,
whose counting, due to the Weyl condition based on $4$-component spinors, is given by the following list
\par
{\em i)} $10$ in the real case ($10=6+4$, in the complex presentation),\par
{\em ii)} $9$  in this real ${\cal R}$ case, \par
{\em iii)} $7$ in this real ${\cal P}$ case,\par
{\em iv)} $6$ in this case, corresponding to the complex hermitian supersymmetry,\par
{\em v)} $4$ in this case, corresponding to the complex holomorphic supersymmetry,\par
{\em vi)} $3$ for a hermitian or holomorphic supersymmetry supplemented by a reality condition,\par
{\em vii)} $1$ for a holomorphic supersymmetry with imaginary ${\cal P}$.\par
The above classes of supersymmetries are present in all cases when the
fermionic generators are realized through complex spinors.\par
Both the $(4,1)$ and the $(3,1)$ spacetimes are not maximal Clifford algebras. The maximal
Clifford algebras associated to them are recovered from the (\ref{maxclifford}) table.
The above list of generalized supersymmetries finds immediate application in the construction
of all possible constrained dynamical systems arising from dimensional reduction of
one given system associated to the maximal Clifford spacetime (examples of such systems
are the particle models admitting tensorial central charges, briefly discussed in the Conclusion). This explains the importance of both the (\ref{maxclifford}) table (for the derivation
of maximal Clifford algebras) and of the above constraints in the classification of generalized supersymmetries.\par
The above results can be extended to any kind of generalized supersymmetries admitting
$n$-component complex spinors (i.e. $2n$ distinct components in the real counting).
In the following table the associated generalized supersymmetries are listed, as well as
the total number of bosonic (real-counting) degrees of freedom.
We have
 {\small
 { {{\begin{eqnarray}&\label{susies}
\begin{tabular}{|l|l|l|}\hline
$i$&$Real\quad supersymmetry$ &$2n^2+n \quad bosonic\quad components$
\\ \hline 

$ii$&$reality \quad on\quad {\cal R}$&$\frac{3}{2}(n^2+n) \quad bosonic\quad components$\\ \hline

$iii$&$reality \quad on\quad {\cal P}$&$\frac{1}{2}(3n^2+n) \quad bosonic\quad components$\\ \hline

$iv$&$complex \quad holomorphic\quad supersymmetry$&$n^2+n \quad bosonic\quad components$\\ \hline

$v$&$complex\quad hermitian$&$
n^2 \quad bosonic\quad components$\\ \hline

$vi$&$complex\quad holomorphic \quad with\quad bosonic\quad reality\quad constraint $&$
\frac{1}{2}(n^2+n) \quad bosonic\quad components$\\ \hline

 $vii$&$complex\quad holomorphic\quad with \quad bosonic\quad imaginary\quad
 constraint
$&$\frac{1}{2}(n^2-n)\quad bosonic\quad components$\\ \hline

\end{tabular}&\nonumber\\&&\end{eqnarray}}} }}  

The constrained generalized supersymmetries labeled by $iii$, $iv$, $v$ and $vi$
admit an equivalent, dual presentation. We discussed this feature in the concrete
example of the Minkowskian $(4,1)$ generalized supersymmetries, but this is a 
general property, valid in any space-time signature. The dual relations are given by
\begin{eqnarray}
iii) &&{\cal P}\, real \leftrightarrow {\cal R}\, imaginary,\nonumber\\ 
iv) &&{\cal R}=0\leftrightarrow {\cal P}, {\cal R}\, both \,real,   \nonumber \\
v) &&{\cal P}=0 \leftrightarrow {\cal P}\, real, {\cal R}\, imaginary,\nonumber\\
vi) &&{\cal P} \, real, {\cal R} =0 \leftrightarrow {\cal R} \, real, {\cal P}=0.
\end{eqnarray}
 
\section{Generalized supersymmetries of the quaternionic spacetimes}

For what concerns the complex and quaternionic cases several tables can be produced
presenting the complete list of the associated constrained supersymmetries. For lack of
space we limit ourselves to reproduce here some selected examples concerning the generalized
supersymmetries supported by quaternionic spacetimes. This provides a reader with an idea 
about the main features of the classification. \par
The first case we present corresponds to the hermitian quaternionic supersymmetry, 
whose fermionic generators
are quaternionic spinors (the corresponding spacetimes 
supporting such spinors and their associated
supersymmetries are given in (\cite{crt1})). In this particular case the 
corresponding table is given by
{ {{\begin{eqnarray}&\label{hh1}
\begin{tabular}{|c|c|c|}\hline
spacetime&bosonic sectors&bosonic components\\ \hline
$D=3$&${ M}_0$& $1$\\ \hline

$D=4$&${ M}_0$& $1$\\ \hline

$D=5$&${ M}_0+{M}_1$&$1+5=6$\\ \hline

$D=6$&${ M}_1$&$6$\\ \hline

$D=7$&${ M}_1+{M}_2$& $7+21=28$\\ \hline

$D=8$&${ M}_2$&$28$\\ \hline

$D=9$&${ M}_2+{M_3}$&$36+84=120$\\ \hline

$D=10$&${ M}_3$&$120$\\ \hline

$D=11$&${ M}_0+{M}_3+{M}_4$&$1+165+330=496$\\ \hline

$D=12$&${M}_0+{ M}_4$&$1+495=496$\\ \hline

$D=13$&${M}_0+{M}_1+{M}_4+{M}_5$&$1+13+715+1287=2016$\\ \hline

\end{tabular}&\end{eqnarray}}} }  
 As an example of holomorphic supersymmetry we produce a table corresponding 
 to the complex holomorphic supersymmetry for quaternionic spacetime, i.e.
 carrying a quaternionic structure, however expressing spinors only through their
 complex structure. This implies that the reality condition on the bosonic
 sector is automatically implemented. Some similarities should be observed between
 the table (\ref{realodd}) and the table below. They correspond however to different cases,
 real versus complex holomorphic supersymmetries, associated to spacetimes with different
 signatures and different number of spinorial components
 (in the complex holomorphic case the number of spinor 
 components are double than in the real case, for 
 $D$-dimensional spacetimes). Their similarities on the other hand have a very deep physical meaning. They imply, e.g., that the complex holomorphic supersymmetry can be used to perform
 the analytic continuation of real supersymmetries to different signatures 
 (the Euclideanized version of the $M$-algebra, see \cite{lt3}, corresponds to the analytical continuation of the real $M$ algebra). 
 We have now
 { {{\begin{eqnarray}&\label{ch11}
\begin{tabular}{|c|c|c|}\hline
spacetime&bosonic sectors&bosonic components\\ \hline
$D=3$&${M}_1$& $3$\\ \hline

$D=4$&${\widetilde{ M}}_2$& $3$\\ \hline

$D=5$&${{M}}_2$&$10$\\ \hline

$D=6$&${\widetilde { M}}_3$&$10$\\ \hline

$D=7$&${ M}_0+{ M}_3$&$1+35=36$\\ \hline

$D=8$&${M}_0+{\widetilde{ M}}_4$& $1+35=36$\\ \hline

$D=9$&${ M}_0+{M}_1+{ M}_4$&$1+9+126=136$\\ \hline

$D=10$&${M}_1+{\widetilde{ M}}_5$&$10+126=136$\\ \hline

$D=11$&${M}_1+{M}_2+{ M}_5$&$11+55+462=528$\\ \hline

$D=12$&${ M}_2+{\widetilde{ M}}_6$&$66+462=528$\\ \hline

$D=13$&${M}_2+{M}_3+{ M}_6$&$78+286+1716=2080$\\ \hline

\end{tabular}&\end{eqnarray}}} }  

The classification of the (full) quaternionic holomorphic supersymmetry, which presents peculiar
features, has been given and discussed in \cite{top}. The results can be summarized as follows

{ {{\begin{eqnarray}&\label{holoquat}
\begin{tabular}{|c|ll|}\hline
$-$ &$D=0,6,7$&\quad mod \quad $8$\\ \hline

${{ M}}_0$& $D=1$&\quad mod \quad $8$\\ \hline

${{ M}}_1$&$D=4,5$&\quad mod \quad $8$\\ \hline

${ M}_0+{ M}_1$&$D=2,3$&\quad mod \quad $8$\\ \hline

\end{tabular}&\end{eqnarray}}} }  

The above results can be interpreted as follows. Quaternionic holomorphic 
supersymmetries
only arise in $D$-dimensional quaternionic space-times, where $D=2,3,4,5\quad mod \quad 8$.
No such supersymmetry exists in $D=0,6,7\quad mod\quad 8$ $D$-dimensional spacetimes. \par In
$D=1 \quad mod\quad 8$ dimensions it only involves a single bosonic
charge and falls into the class of quaternionic supersymmetric quantum mechanics, rather than
supersymmetric relativistic theories.
Finally, this supersymmetry algebra only admits at most a scalar bosonic central charge, found
in $D$-dimensional quaternionic spacetimes for $D=2,3\quad mod\quad 8$.\par
It must be said that so far no dynamical system 
supporting such a supersymmetry has been investigated. 

\section{Conclusions.}

This paper was devoted to perform a division algebra classification of the generalized supersymmetries. Besides the notion of hermitian (complex and quaternionic) and holomorphic (complex and quaternionic) supersymmetries, already presented in \cite{top}, 
a further distinction of division-algebra constrained generalized supersymmetries, given by table (\ref{susies}), has been presented. This set of constrained supersymmetries corresponds to
certain classes of division algebra constraints that can be consistently imposed (e.g., a reality condition on the bosonic sector of complex holomorphic supersymmetries ). 
The sets of constraints can even be combined together, as discussed in Section {\bf 6}.\par
Another issue that we have here clarified consists in the explicit construction, see table
(\ref{maxclifford}), of the non-maximal Clifford algebras and their
associated spinors, in terms of their associated maximal Clifford algebras.
The two main new results here presented allow to classify and put in a single framework
(via dimensional reduction), showing their web of inter-related dualities, a
whole class of generalized supersymmetries.They can be combined to produce, on a physical side,
the largest ``oxydized" dynamical system which can be regarded as the generator of
all reduced and constrained lowest dimensional models.\par
Some of the mathematical issues here discussed have already been employed to, e.g.,
performing the analytic continuation of the $M$ algebra \cite{lt3} (it corresponds to
an eleven-dimensional complex holomorphic supersymmetry and in \cite{top} 
it was further shown that the same algebra
also admits a $12$-dimensional Euclidean presentation in terms of Weyl-projected spinors). These two examples
of Euclidean supersymmetries can find application in the functional integral formulation of higher-dimensional
supersymmetric models. \par
There is an interesting class of models which nicely fits in the framework here described and is currently under
intense investigation. It is the class of superparticle models, introduced at first in \cite{rs} and later studied in \cite{bl},
whose bosonic coordinates correspond to tensorial central charges. It was shown in \cite{blps} that a $4$-dimensional
theory of this kind leads to a tower of massless higher spin states, concretely implementing a Fronsdal's
proposal \cite{fro} of introducing bosonic tensorial coordinates to describe massless higher spin theories (admitting helicity
states greater than two). This is an active area of research, the main motivation being the  investigation of the
tensionless limit of superstring theory, corresponding to a tower of higher helicity massless particles
(see e.g. \cite{sor}).
\par
In a somehow ``orthogonal" direction, a class of theories which can be investigated in the present framework
is the class of supersymmetric extensions of Chern-Simon supergravities in higher dimensions, requiring as a basic
ingredient a Lie superalgebra admitting a Casimir of appropriate order, see e.g. \cite{htz}.

\end{document}